\documentclass[ amsmath,amssymb,notitlepage,preprint,longbibliography]{revtex4-1}

\usepackage{graphicx}
\usepackage{dcolumn}
\usepackage{bm}
\usepackage[caption=false]{subfig}
\usepackage{txfonts}
\usepackage{gensymb}
\usepackage{color}
\usepackage{soul}
\usepackage{ulem}

\begin{document}

\preprint{APS/123-QED}

\title{Numerical study of the McIntyre instability around {Gaussian} floating vortices { in thermal wind balance}}

\author{M. Le Bars}
\affiliation{CNRS, Aix Marseille Univ, Centrale Marseille, IRPHE, Marseille, France}

\date{\today}

\begin{abstract}

The visco-diffusive McIntyre instability \cite{mcintyre1970diffusive} has been suggested as a possible source for density layer formation around laboratory and oceanic vortices. This suggestion is here quantitatively addressed using idealised, axisymmetric, numerical simulations of a simple Gaussian-like vortex in thermal wind balance, floating in a rotating, stratified flow{. Numerical simulations are} complemented by a local stability analysis derived from the seminal study \cite{mcintyre1970diffusive}. It is confirmed that the McIntyre instability is responsible for the layering observed around laboratory vortices, but its relevance for explaining layering around meddies remains doubtful.

\end{abstract}

\maketitle

\section{\label{sec:intro}Introduction}

Long-lived, floating vortices are common features of geophysical and astrophysical flows subject to the combined effects of rotation and stable stratification. They are for instance expected in protoplanetary disks  \cite{barranco2005three}, where they could initiate the formation of planets by concentrating dust particles \cite{barge1995did}. Jupiter's Great Red Spot (GRS) is supposed to be the signature, within the cloud layer, of a large anticyclone floating in the turbulent atmosphere \cite{marcus1993jupiter,lemasquerier2020remote}. And the most accessible floating vortices in Nature are undoubtedly the oceanic ones, in particular the so-called meddies that have been the focus of many campaigns \cite{bashmachnikov2015properties} since their discovery off the Bahamas in the late 70's \cite{mcdowell1978mediterranean}.

\begin{figure}
	\centering
	\includegraphics[width=\linewidth]{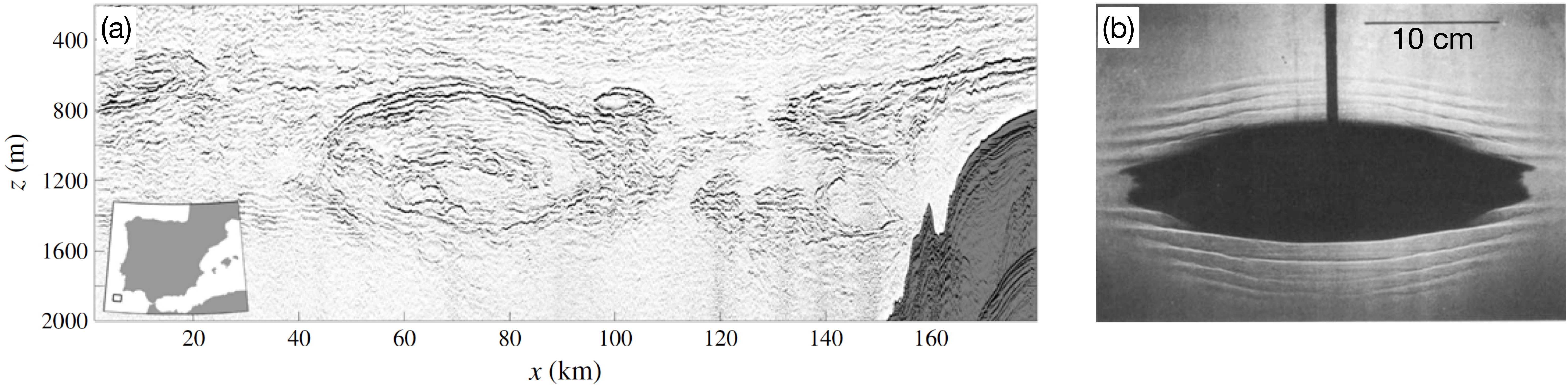}
	\caption{(a) Reproduced with permission from \cite{hua2013layering}: layering revealed by seismic reflection around a meddy in the Gulf of Cadiz (see location in the inset). The displayed quantity is the product of the fluid density by the sound speed (the stronger, the darker) in a vertical cross section of the water column, with $z=0$ corresponding to the sea surface. Layers are $10-100$~m thick and extend quasi-horizontally over tens of kilometers. (b) Reproduced with permission from \cite{griffiths1981stability}: layering revealed by side-view shadowgraphy around a laboratory floating vortex produced by a constant, small flux of dyed fluid in a rotating, linearly stratified ambient with Coriolis frequency $ f = 1.1$~rad/s and buoyancy frequency $ N = 0.66$~rad/s. }
	\label{fig:intro}
\end{figure}

Meddies, a short name for Mediterranean eddies, are isolated lenses of warm and salty Mediterranean water floating in the deep of the Atlantic ocean. They form close to the Gibraltar Strait, where the dense Mediterranean water first flows down the Atlantic continental slope, reaches a level of neutral buoyancy, spreads horizontally, and finally organises as an anticyclone through the action of the Coriolis force. As listed by \cite{richardson2000census}, meddies are typically $0.5 - 1$~km high, $50 - 100$~km large, and can persist for up to 5 years, crossing the Atlantic until they crash on the opposite coast. Meddies share with other floating vortices like the GRS, a pancake ellipsoidal shape (see figure \ref{fig:intro}), a rather strong anticyclonic motion, and a surprising longevity. 

Those specificities have been the subject of dedicated, idealized, model studies combining analytical, experimental and numerical approaches \cite[see e.g.][]{griffiths1981stability,gill1981homogeneous,hedstrom1988experimental,hassanzadeh2012universal,aubert2012universal,ungarish2015coupling,facchini2016lifetime,burin2020instabilities}. In short, the ellipsoidal shape of a floating vortex is due to the geostrophic hydrostatic equilibrium taking place at zeroth order in dissipation and first order in Rossby number, that fixes its aspect ratio
\begin{equation} \label{eq:aspect}
\alpha=\frac{H}{L}=\sqrt{\frac{-Ro}{N^2-N_c^2}}f,
\end{equation}  
where $H$ and $L$ are its mid-height and radius, $f$ and $N$ the Coriolis and ambient buoyancy frequencies, $N_c$ the buoyancy frequency in the vortex core, and $Ro$ the Rossby number equal to the ratio of the angular velocity at the vortex center divided by $f$ \cite{hassanzadeh2012universal,aubert2012universal}. For stability, a sub-stratified vortex compared to the ambient with $N_c < N$, such as the meddies and GRS, requires $Ro<0$, hence an anticyclonic motion. Then, at first order in dissipation, the longterm, slow temporal evolution of a floating vortex is related to the diffusion of the main equilibrium state, which generates an internal recirculation through the combined effect of rotation and stratification \cite{facchini2016lifetime}. In the limit of small aspect ratio $\alpha$ and large Schmidt number Sc (ratio of viscosity $\nu$ to the stratifying-agent diffusivity $D$), this secondary motion leads to a typical laminar duration
\begin{equation} \label{eq:time}
\tau=\frac{L^2}{\nu}\frac{f^2}{N^2},
\end{equation}  
which replaces the usual viscous dissipation timescale $\tau_{usual}={H^2}/{\nu}$. Scaling laws (\ref{eq:aspect})--(\ref{eq:time}), validated in idealised laboratory models, provide the basis for interpreting oceanic measurements and for predicting e.g. the unknown depth of the GRS or the duration of a given meddy. Natural flows are of course complexified by additional effects like background turbulence, large-scale currents, radiative damping, compressibility, etc. Idealized models nevertheless perform quite well \cite[e.g.][]{aubert2012universal,facchini2016lifetime,lemasquerier2020remote}.

Another surprising feature of meddies has been revealed by geoseismic observations (see e.g. \cite{biescas2008imaging} and figure \ref{fig:intro}(a)): the presence, in all their vicinity, of sharp density interfaces separated by thin elongated quasi-horizontal mixed layers with a typical thickness of 10 to 100~m. These layers  represent, in the oceans, the physical manifestation of a bulk interior route to dissipation of the
energy injected at the planetary scale \cite{hua2013layering}. In addition to boundary flows and internal gravity waves, layering could hence participate in closing the ocean energy budget \cite{muller2005routes}. Yet the origin of layering is still debated. Since Mediterranean water is both saltier and warmer than the Atlantic water, double-diffusive processes, involving competitive diffusion of heat and salt, are obvious contributors to layer formation \cite{ruddick1988mixing,radko2017life}. However, quasi-geostrophic simulations neglecting salinity variations also exhibit strong layering around meddies, then explained by the non-linear development of a baroclinic, critical layer
instability \cite{hua2013layering}. In the laboratory, layering is clearly seen in the seminal experiment by Griffiths \& Linden \cite{griffiths1981stability}, where dyed, medium-salty water is slowly and continuously poured at its neutral level of buoyancy into a rotating, linearly stratified ambient (see figure \ref{fig:intro}(b) for a reproduction of their famous photo). This experiment uses salt only (no thermal effect), hence precluding a double-diffusive origin for the observed layering; and the observed pattern is a priori axisymmetric, hence precluding a critical layer origin. The authors then attributed this layering to the visco-diffusive, {McIntyre} instability \cite{mcintyre1970diffusive}, yet with no systematic study nor definitive proof, while possible sources of instability at the laboratory scale are numerous \cite[see e.g.][]{yim2016stability}. 

The present study complements Griffiths \& Linden's original work \cite{griffiths1981stability} by providing a systematic analysis of layering formation around floating vortices in idealised, axisymmetric, numerical simulations. The purpose is threefold: first, to undoubtedly confirm that the observed layering is indeed due to the McIntyre instability \cite{mcintyre1970diffusive}; then to evaluate how well the observed flows compare with McIntyre's predictions in terms of growth rate and most unstable lengthscale, as derived from his generic local analysis; and finally to explore the possible relevance of this visco-diffusive mechanism for layering formation around real meddies, acknowledging that the observed natural pattern may in fact come from the superposition of several effects. 

The paper is organized as follows. Section \ref{sec:local} presents a rapid overview of McIntyre's historical approach and of its previous experimental validations. The numerical model is then described in section \ref{sec:num}. Systematic numerical results are presented and analysed in section \ref{sec:syst}. Section \ref{sec:app} discusses possible application to real meddies. Finally, conclusions and directions for future work are provided in section \ref{sec:conclusion}.

\section{A short overview of the McIntyre local approach and of its previous experimental validation} \label{sec:local}

The main results from the historical, analytical study of McIntyre first published in 1970, are briefly presented here. Interested readers should refer to the original paper \cite{mcintyre1970diffusive} as well as to its more recent reformulation in \cite{munro2010instabilities} for additional details. Note for completeness that the generic mechanism described by McIntyre was discovered shortly before 1970 in the astrophysical community in the specific limit $\mbox{Sc} \ll 1$ \cite{goldreich1967differential,fricke1968instabilitat}: it is then referred to as the Goldreich-Schubert-Fricke (GSF) instability \cite[see details in][]{barker2020angular}. { In the fluid mechanics and oceanography litterature, McIntyre instability is also sometimes called ``diffusive instability'', ``viscous overturning instability'',  ``viscous-diffusive instability'', and ``viscous-diffusive overturning''}.

McIntyre performed a local, linear stability analysis of a generic, axisymmetric base flow in thermal wind balance
\begin{equation} \label{eq:baseflow}
\frac{g}{\rho_0} \tilde{\rho'}_r= -f \tilde{v}_z,
\end{equation}  
where ${\rho_0}$ is the reference density, $g$ the acceleration of gravity, $\tilde{\rho'}$ the perturbation density compared to the hydrostatic profile including the background linear stratification $N$, $\tilde{v}$ the azimuthal velocity in the frame rotating at $f/2$, and where subscripts stand for partial derivatives considering the usual cylindrical coordinates. { Munro et al.  \cite{munro2010instabilities} extended the study to gradient wind balanced flows without reporting any significant change. }McIntyre looked at the evolution of an axisymmetric, plane wave perturbation, assuming scale separation between the small-scale perturbation and the large-scale possible variations of the base flow, neglecting curvature effects {by locally considering the cylindrical coordinates as Cartesian ones}, but keeping diffusion terms. He derived the dimensionless dispersion relation
\begin{equation} \label{eq:dispersion}
\omega^3+k^2 \left( 2+\frac{1}{\mbox{Sc}}\right) \omega^2+ \left( G+I+k^4  \left( 1+\frac{2}{\mbox{Sc}}\right) \right) \omega +k^2 \left( G+\frac{I+k^4}{\mbox{Sc}}\right)=0.
\end{equation}  
Here, $k$ is the wavevector norm non-dimensionalized by the viscous lengthscale 
\begin{equation}  \label{eq:scaleL}
\delta = \left( \frac{\nu^2 \rho_0}{g|\tilde{\rho'}_r|} \right)^{1/4}, 
\end{equation}  
and $\omega$ is the growth rate non-dimensionalized by the timescale 
\begin{equation} \label{eq:scaleT}
\tau =\left( \frac{\rho_0}{g|\tilde{\rho'}_r|} \right)^{1/2}.
\end{equation}  
$G$ and $I$ are two functions that depend on the wavevector angle with the horizontal $\phi$, as well as on the angles $\Gamma$ and $\Theta$ that the lines of constant circulation and the isopycnals of the complete base flow (including global rotation and linear stratification) respectively make with the vertical:
\begin{equation} 
G= \frac{\cos{\phi}}{\cos{\Theta}}\sin({\Theta-\phi}),
\end{equation} 
\begin{equation}  \label{eq:I}
I= \frac{\sin{\phi}}{\sin{\Gamma}}\sin({\phi-\Gamma}).
\end{equation} 
Threshold for instability is given by the simple relation
\begin{equation}  \label{eq:threshold}
\frac{\tan{\Theta}}{\tan{\Gamma}} < \frac{\mbox{Sc}}{4}\left(1+\frac{1}{\mbox{Sc}} \right)^2,
\end{equation} 
which is especially illuminating regarding the origin of the instability.  Indeed,  { the inviscid, non-diffusive limit of the dispersion relation (\ref{eq:dispersion}), after replacing (\ref{eq:scaleL}) by a new lengthscale independent of viscosity, gives
\begin{equation} \label{eq:dispersioninviscid}
\omega^3+ \left( G+I \right) \omega =0,
\end{equation}  
hence the threshold 
\begin{equation}  \label{eq:thresholdinviscid}
\frac{\tan{\Theta}}{\tan{\Gamma}} < 1.
\end{equation} 
This threshold characterises ``classical'' inviscid instability mechanisms, including inertial, symmetric,  and centrifugal instabilities. It is noteworthy that accounting for viscosity and diffusion, which a priori hamper classical instabilities,  }
the right-hand {side} of equation (\ref{eq:threshold}) admits a minimum also equal to $1$ for $\mbox{Sc} =1$.  But considering dissipation with $\mbox{Sc} \neq 1$, equation (\ref{eq:threshold}) becomes less restrictive, meaning that dissipation with $\mbox{Sc} \neq 1$ actually promotes instability: for a given balanced, inviscidly stable flow with ${\tan{\Theta}}/{\tan{\Gamma}} >1$, there exist a critical Schmidt number $\mbox{Sc}_{cr,>1} > 1$ above which and a critical Schmidt number  $\mbox{Sc}_{cr,<1} < 1$ below which the flow becomes unstable. 

The underlying mechanism of this visco-diffusive,  {McIntyre} instability has a lot in common with the classical double-diffusive instability, which requires a {Lewis} number different than $1$ \cite[see e.g.][]{radko2013double}. Let's consider a balanced flow and move a fluid parcel away from its equilibrium position. The fluid parcel re-equilibrates with its new environment by diffusion. Considering for instance $\mbox{Sc} >1 $ as for salty water, its momentum equilibrates faster than its buoyancy: hence for a carefully chosen initial displacement direction, one may expect this displaced fluid parcel to become gravitationally unstable once in momentum balance with its new environment. The angle selection of the McIntyre instability is imbedded in the dispersion relation (\ref{eq:dispersion}), where the growth rate depends on $G$ and $I$, hence on the relative values of the wavevector angle $\phi$ vs. the base flow angles $\Gamma$ and $\Theta$  \cite[see a detailed discussion { and illuminating illustrations} in][]{ruddick1992intrusive}. One can also notice that the only dimensionless parameter that appears in the dispersion relation (\ref{eq:dispersion}), for a given orientation of the base flow, is the Schmidt number: $\mbox{Sc}$ solely determines the threshold as well as the growth rate and wavelength of the most unstable mode, adimensionalised by the specific choice of length and time scales (\ref{eq:scaleL})--(\ref{eq:scaleT}). Coming back to a more classical adimensionalisation using the background rotation rate $f/2$ and the typical base flow lengthscale $L$, the system of course also depends, as usual, on the Ekman number $\mbox{Ek}=\nu/(L^2 f/2 )$, the base flow Rossby number, and the ratio $f/N$. From (\ref{eq:scaleL})--(\ref{eq:scaleT}), one can then straightforwardly predict that, with other parameters being constant, the most unstable wavelength scales as $\mbox{Ek}^{1/2}$ and the growth rate is independent on $\mbox{Ek}$, a {counter-intuitive} result for such a viscous instability.

McIntyre instability was first observed experimentally using shadowgraphy by \cite{baker1971density} in a stratified spin-up experiment, where an horizontal disk is set in differential rotation within a rotating tank of linearly stratified, salt water. The same set-up was then re-investigated with improved metrology, including density measurements by conductivity probes \cite{calman1977experiments} and particle image velocimetry \cite{munro2010instabilities}. The instability manifests as regularly spaced density layers superimposed on the background density profile. Systematically changing the ratio $f/N$, the Rossby number via the imposed differential rotation, as well as the Schmidt number via changes of the (uniform) fluid temperature, convincing quantitative agreement with the linear stability results was shown for the stability threshold, as well as for the wavelength and growth rate of the most unstable mode \cite{munro2010instabilities}. 

The situation is less clear for layering around a  floating vortex. Griffiths \& Linden \cite{griffiths1981stability} (see also figure \ref{fig:intro}b) report a clear layering, but for one single set of parameters only. \cite{hedstrom1988experimental} also reports layering above a threshold in quantitative agreement with the linear stability analysis, but again no systematic exploration of the parameter space is provided. Actually the poor control and restricted range of approachable parameters in laboratory floating vortices render a complete experimental study very difficult. As described in \cite{aubert2013formes}, only vortices generated by continuous injection of mid-density fluid clearly show layering, but then the Rossby number can hardly be adjusted: it is the result of the injection process combined with the Coriolis force. Additionally, the base flow and vortex size continuously evolve in time because of the injection. Besides, the ratio $f/N$ in such experiments is constrained by the limited range of buoyancy frequency accessible with salt water, by the necessity to impose a large enough rotation rate to limit the predominance of viscous effects (i.e. small enough $\mbox{Ek}$), and by the upper limit of the turntable rotation rate.  And finally, up to now, only layering around salty water floating vortices at ambient temperature has been reported, hence fixing the central control parameter $\mbox{Sc} \simeq 700$. 
For definitively proving the McIntyre origin of the layering around laboratory floating vortices, we thus turn towards complementary numerical simulations that allow to explore a large range in all dimensionless control parameters, namely $\mbox{Sc}, \mbox{Ek}, f/N$ and $Ro$.

\section{Numerical method} \label{sec:num}

Let us consider a floating vortex {with a Gaussian azimuthal velocity profile and zero radial and vertical velocities}, as frequently used to approximate isolated oceanic vortices \cite[e.g.][]{radko2017life} as well as isolated experimental vortices produced by injection \cite[e.g.][]{beckers2001dynamics,de2017laboratory}, including floating vortices \cite{facchini2016lifetime}.  Adimensionalisation is classically performed using the background rotation rate $f/2$ and the vortex radius $L$, and for convenience the density scale $\rho_0  \frac{N^2 L}{g}$.  We assume that this base flow is maintained in thermal wind balance by some ``bulk forces'' which compensate the viscous damping, the diffusion of the stratifying agent, and the centrifugal force in Navier-Stokes equations. Then, following the thermal wind balance (\ref{eq:baseflow}) and hence, the equilibrium aspect ratio $\alpha$ given by (\ref{eq:aspect}), the base state writes
\begin{equation} \label{eq:gaussprofile}
\left(\tilde{v},\tilde{\rho'}\right) = (2 {Ro} \times r, z) \exp{\left(-r^2-\frac{z^2}{\alpha^2}\right)},
\end{equation}  
focusing here on $N_c=0$, which is relevant for laboratory vortices during and shortly after injection (see {e.g. \cite{stuart2011geostrophic}, but also }discussion about longer term evolution in \cite[][]{lemasquerier2020remote}). Considering the generic case with $N_c \neq0$ would simply add a prefactor $1-(N_c/N)^2$ to the perturbation density profile, hence would add one more dimensionless parameter to the already 4-dimensions parameter space to be explored, but with no fundamental change on the underlying physical process for instability.
{ Accounting for finite Rossby numbers effects,  one could also choose a base flow in gradient wind rather than in thermal wind balance.  Analytical formulae for the base velocity and/or density profiles would then be  more complex than (\ref{eq:gaussprofile}) \cite[see e.g.][]{hassanzadeh2012universal}. They would anyhow remain approximation of the real base state for both meddies and laboratory vortices.  For simplicity and consistency with McIntyre's historical study, I thus focus here on the thermal wind balanced base flow (\ref{eq:gaussprofile}).}

Let's thus assume that it is imposed and maintained externally (mimicking the slow, continuous injection in the lab, but with no volume change), and solve for the axisymmetric perturbation flow around it. The velocity, density and pressure fields $(u,v,w,\rho,p)$ are then solutions of the following set of dimensionless Navier-Stokes equations in the Boussinesq approximation, written in cylindrical coordinates
\begin{equation} \label{eq:U}
\partial_t u + u \partial_r u + w \partial_z u - \frac{v^2}{r} - \frac{2v\tilde{v}}{r} - 2v = -\partial_r p + \mbox{Ek} \left(\Delta_{a}  u -\frac{u}{r^2} \right),
\end{equation}  
\begin{equation} \label{eq:V}
\partial_t v + u \partial_r v + w \partial_z v + u \partial_r \tilde{v} + w \partial_z \tilde{v} + \frac{uv}{r} + \frac{u  \tilde{v}}{r} + 2u = \mbox{Ek} \left(\Delta_{a}  v -\frac{v}{r^2} \right),
\end{equation}  
\begin{equation} \label{eq:W}
\partial_t w + u \partial_r w + w \partial_z w = -\partial_z p - \left( \frac{2N}{f}\right)^2 \rho+ \mbox{Ek} \Delta_{a}  w,
\end{equation}  
\begin{equation} \label{eq:rho}
\partial_t \rho + u \partial_r \rho + w \partial_z \rho + u \partial_r \tilde{\rho'} + w \partial_z \tilde{\rho'} -w = \frac{\mbox{Ek}}{\mbox{Sc}} \Delta_{a}  \rho,
\end{equation}  
\begin{equation} \label{eq:div}
\frac{1}{r}\partial_r(ru)+\partial_z w =0,
\end{equation}  
with $\Delta_{a} = 1/r\partial_r(r\partial_r) + \partial_{z^2}$.
Equations (\ref{eq:U})--(\ref{eq:div}) are solved using the commercial code COMSOL Multiphysics, based on the finite element method. The domain has the same ellipsoidal shape as the background Gaussian vortex, but with a radius $2.5$. 
The outer boundary conditions are no flux and free slip. It has been checked that this choice of domain size and boundary conditions does not significantly influence the obtained results. Assuming axisymmetry and equatorial symmetry allows to reduce the domain to its upper right quarter. The code uses a triangular mesh, strongly refined within and close to the vortex (i.e. up to radius $1.5$), and standard Lagrange elements which are quadratic for the pressure and cubic for the density and velocity fields. The total number of degrees of freedom is at minimum $891224$: it depends on the dimensionless parameters via the aspect ratio $\alpha$ that changes the domain size, and it is also adjusted to ensure a minimum of 8 elements per unstable wavelength. Grid convergence tests have been systematically performed, especially for large values of the Schmidt number and/or small values of the Ekman number. The time-dependent solver uses the generalized alpha method, with similar properties to a second-order backward difference method. The sparse direct linear solver is Pardiso. No stabilization technique is used. 

\begin{figure}
	\centering
	\includegraphics[width=\linewidth]{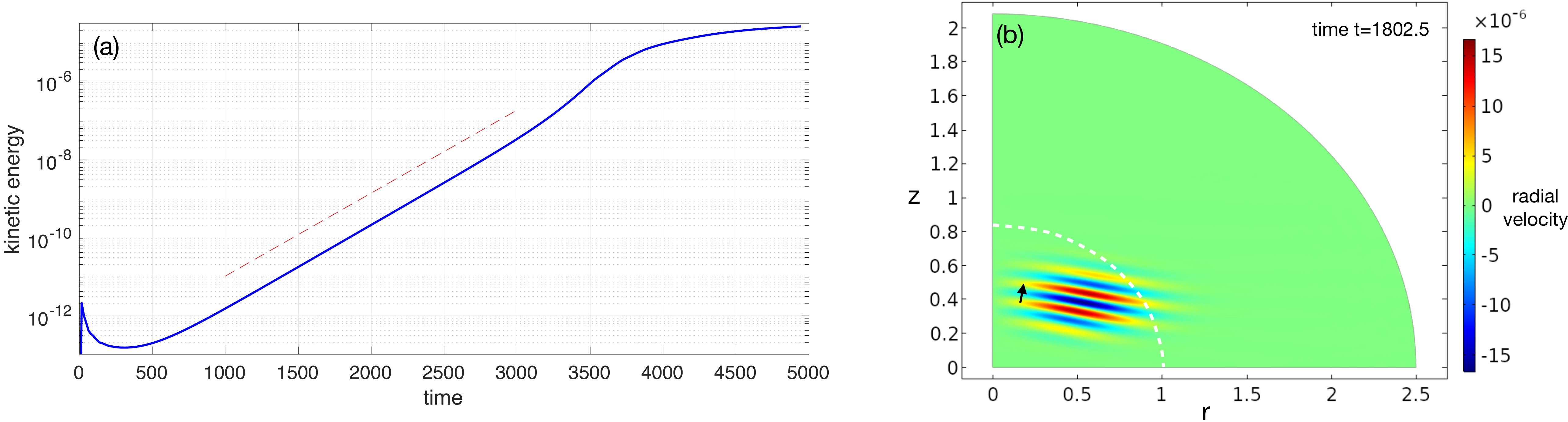}
	\caption{Numerical results obtained from the reference computation with $\mbox{Sc} = 200$, $\mbox{Ek}=3.5\times 10^{-4}$, $\mbox{Ro}=-0.25$, and $N/f=0.6$. (a) shows the temporal evolution of the kinetic energy (blue) and its exponential bestfit (dashed red), with a growth rate $2\omega=4.78 (\pm 0.07) \times 10^{-3}$ (the factor $2$ comes from considering here the kinetic energy). (b) shows a snapshot at time $t=1802.5$ of the radial velocity field. The black arrow shows the determined wavevector by wavelet analysis over the exponential growth phase, with $2\pi/k=0.113 \pm 0.002$ and $\phi=80.2^o (\pm 0.7) $. The white dashed line shows the edge of the Gaussian vortex defined here as $r^2+(z/\alpha)^2=1$.}
	\label{fig:ex}
\end{figure}

A small amplitude noise, typically $10^{-4}$ the background amplitude, is added to the density field as the initial condition, and the code is then run for 250 to 5000 rotations, i.e. until the beginning of the saturation phase of the instability or until dissipation of the initial noise. The reference computation considers $\mbox{Sc} = 200$, $\mbox{Ek}=3.5\times 10^{-4}$, $\mbox{Ro}=-0.25$, and $N/f=0.6$. The chosen values of $\mbox{Ek}$ and $N/f$ correspond to the experiment of Griffiths \& Linden \cite{griffiths1981stability}: figure \ref{fig:intro}(b) allows to unambiguously measure the vortex depth $2H$, and then to determine its radius $L$ via the provided injection rate $Q=1.3$~cm$^3$/s and time $t=780$~s. $\mbox{Ro}$ is not accessible in their set-up, but applying the equilibrium aspect ratio law  (\ref{eq:aspect}) with $N_c=0$ gives $\mbox{Ro}=-0.15$. To consider the Schmidt molecular value for salt water $\mbox{Sc} = 700$ is computationally demanding: the Schmidt number has thus been purposefully decreased to $200$ for the reference case for easier numerical convergence, and the Rossby number has accordingly been slightly decreased to $\mbox{Ro}=-0.25$ for exciting significant instability. 

From this reference case, the influence of each dimensionless parameter is successively explored within the range  $50 \leq \mbox{Sc} \leq 700$, $3.5\times 10^{-6} \leq \mbox{Ek} \leq 3.2\times 10^{-3}$, $-0.45 \leq \mbox{Ro} \leq -0.15$, and $0.15 \leq N/f \leq 2.4$. Note that we focus here on laboratory salt vortices, hence on $\mbox{Sc} >1$ cases only. For each of the 31 simulations used below, the temporal evolution of the total kinetic energy is first computed, and for the unstable cases, an exponential fit during the initial growth determines the growth rate (see figure \ref{fig:ex}(a)). The most unstable mode during the exponential growth is then determined by performing a wavelet analysis of the radial velocity signal using Matlab ``cwt'' function with Morlet wavelets. The radial velocity field is analysed each 2.5 rotations, then all results are averaged and the most energetic wavevector is selected: an example is shown in figure \ref{fig:ex}(b). Uncertainties are determined by slightly changing the ranges in time and space over which the bestfit and wavelet analysis are performed. Note that in the simulations, layering first appears within the vortex around the most unstable point located in the vicinity of $(r_0=1/2,z_0=\alpha/2)$. Layers then rapidly extend horizontally and vertically, leading to a pattern similar to the laboratory and oceanic observations.

For comparison with the McIntyre linear analysis, the dispersion relation (\ref{eq:dispersion}) is also solved for each case, looking for the largest growth rate $\omega$ over all possible locations $(r_0,z_0)$ within and around the vortex and over all possible wavectors $(k,\phi)$. For better consistency with the numerical system given by (\ref{eq:U})--(\ref{eq:div}), we make one correction to the historical study of McIntyre: the {centrifugal} terms $-{2v\tilde{v}(r_0,z_0)} / {r_0}$ in (\ref{eq:U}) and ${u\tilde{v}(r_0,z_0)} / {r_0}$ in (\ref{eq:V}) are also accounted for in our linear study. Using McIntyre notation and formalism, a straightforward derivation shows that this simply adds a factor $1+2{\tilde{v}(r_0,z_0)} / {r_0f}$ to the definition of the function I given by equation (\ref{eq:I}). The threshold (\ref{eq:threshold}) then accordingly writes
\begin{equation} 
\frac{\tan{\Theta}}{\tan{\Gamma}} < \frac{\mbox{Sc}}{4}\left(1+\frac{1+{2\tilde{v}(r_0,z_0)}/{r_0f}}{\mbox{Sc}} \right)^2.
\end{equation} 
{This} is more appropriate to the Gaussian vortices studied here, where the corrective term $\tilde{v}(r_0,z_0) / {r_0}$ is of the same order as the gradient terms $\left( \partial_r\tilde{v}(r_0,z_0),\partial_z\tilde{v}(r_0,z_0) \right)$ driving the instability.

\section{Numerical results and comparison with the local analysis} \label{sec:syst}

\begin{figure}
	\centering
	\includegraphics[width=.7\linewidth]{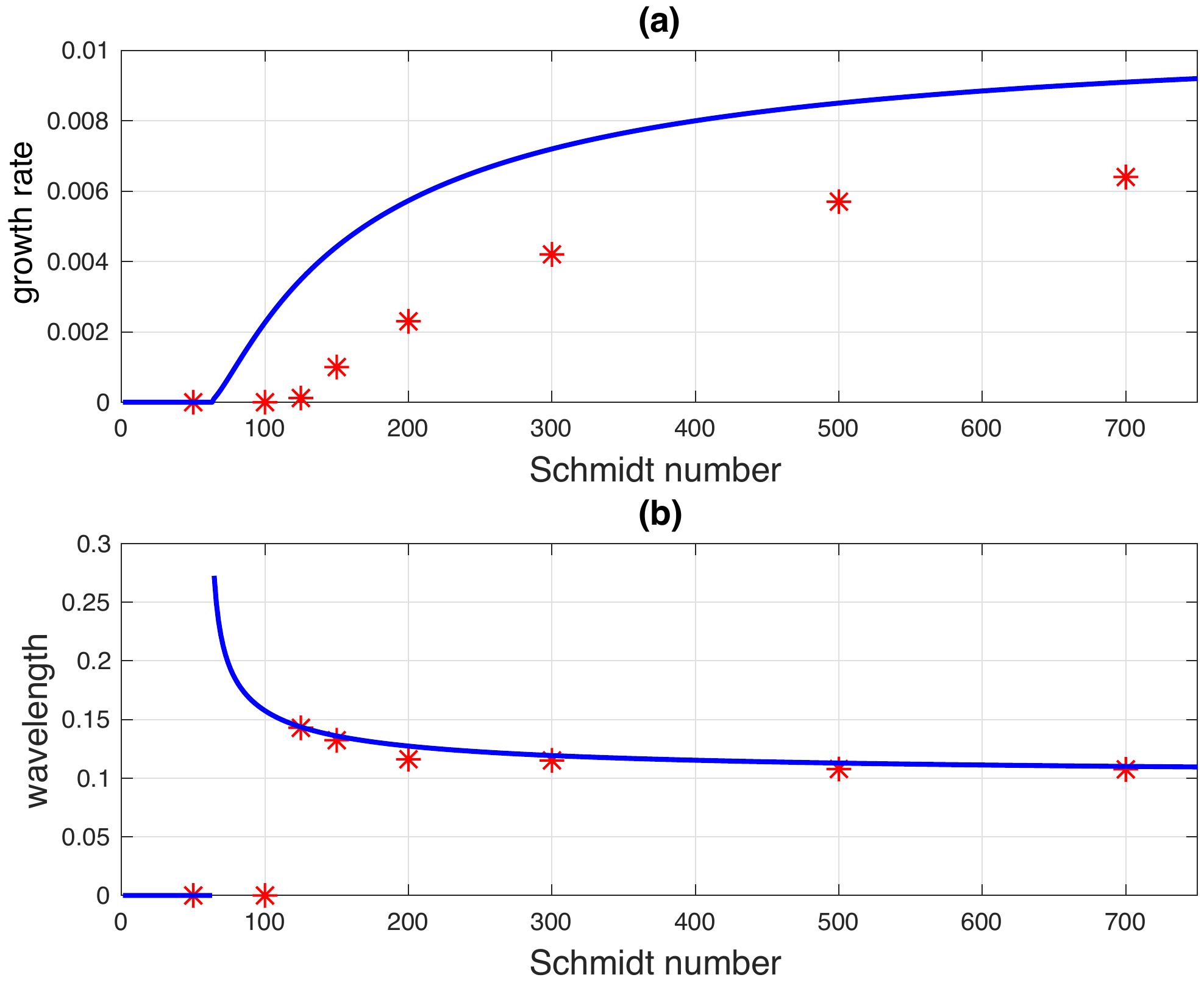}
	\caption{Systematic study of the growth rate (a) and of the most unstable mode wavelength (b) as a function of the Schmidt number $\mbox{Sc}$ for $\mbox{Ek}=3.5\times 10^{-4}$, $\mbox{Ro}=-0.25$, and $N/f=0.6$. The blue curves show predictions from the McIntyre linear theory {accounting for the centrifugal terms} for $\mbox{Sc}\geq 1$, and red stars show numerical results. Symbol size is larger than estimated uncertainties.}
	\label{fig:Sc}
\end{figure}

Let's first perform a systematic study in Schmidt number, in order to unambiguously confirm the visco-diffusive origin of the observed instability. Results are shown in figure \ref{fig:Sc}. The existence of a critical Schmidt number $\mbox{Sc}_{cr,>1} = 121.5 \pm 0.5$ for the numerical simulations, as well as the {good} agreement between the measured and predicted wavelengths, make two convincing arguments for the McIntyre mechanism. The increases of the predicted and measured growth rates with the Schmidt number share a similar functional form characteristic of a supercritical instability. However the theoretical prediction systematically overestimates instability, with e.g. a threshold $\mbox{Sc}_{cr,>1} = 63.9$. It is argued that this quantitative difference is due to the limited scale separation in the configuration studied numerically, where the most unstable wavelength is of order $0.1$ or even larger close to threshold (see figure \ref{fig:Sc}(b)): this absence of significant scale separation limits the quantitative validity of the local approach, even if the underlying physical process is valid. The {good} agreement between the measured and predicted wavelengths is then surprising.

\begin{figure}
	\centering
	\includegraphics[width=.7\linewidth]{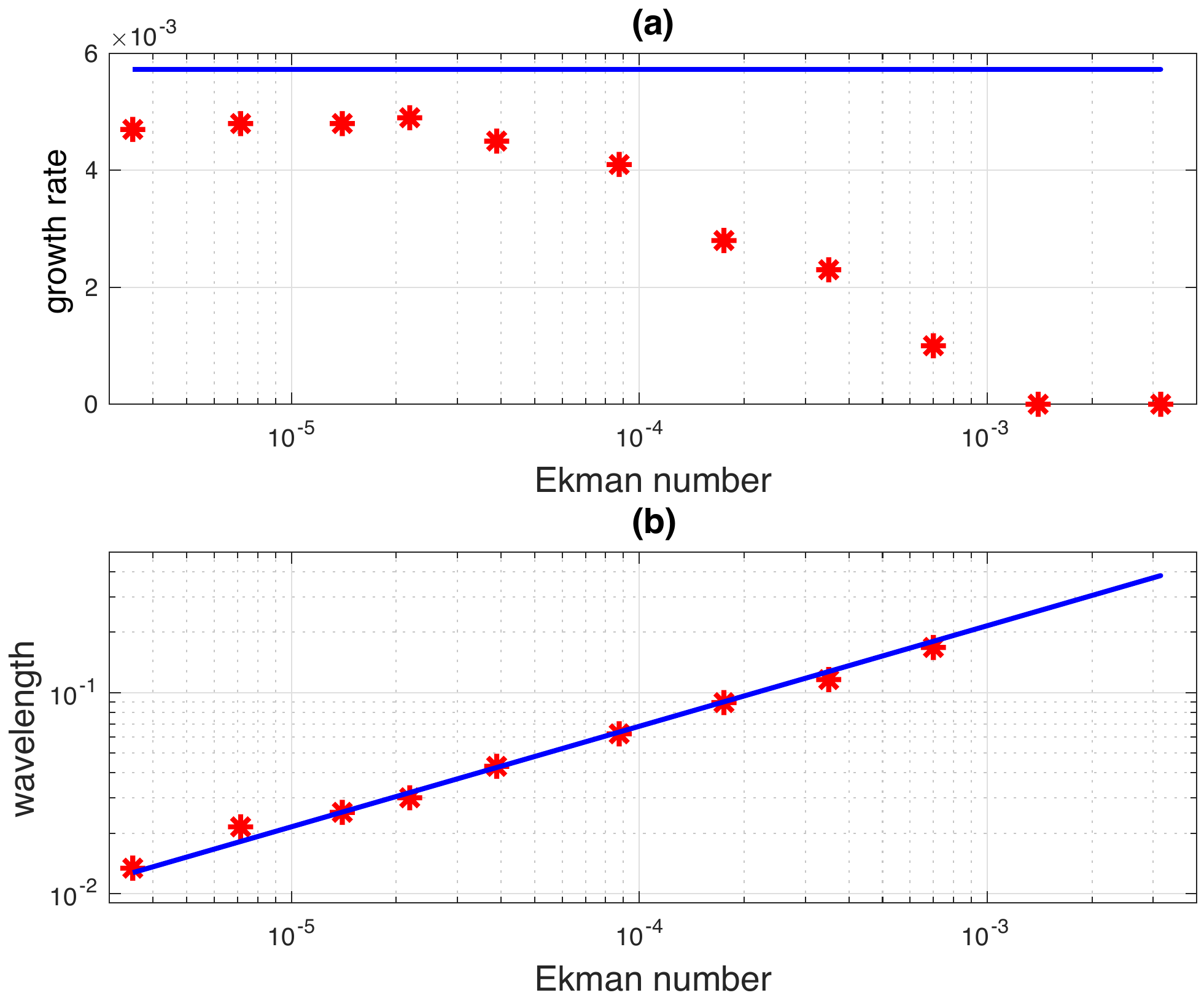}
	\caption{Systematic study of the growth rate (a) and of the most unstable mode wavelength (b) as a function of the Ekman number $\mbox{Ek}$ for $\mbox{Sc}=200$, $\mbox{Ro}=-0.25$, and $N/f=0.6$. The blue curves show predictions from the McIntyre linear theory {accounting for the centrifugal terms}, and red stars show numerical results. Symbol size is larger than estimated uncertainties.}
	\label{fig:Ek}
\end{figure}

To further prove the importance of scale separation, figure \ref{fig:Ek} shows the systematic study as a function of the Ekman number. Again, the measured and predicted wavelengths {closely} agree, and recover the theoretically expected scaling in $\mbox{Ek}^{1/2}$ (figure \ref{fig:Ek}(b)). Hence, scale separation is better fullfilled with smaller unstable modes when decreasing $\mbox{Ek}$: the measured growth rate jointly increases and saturates closer to the predicted value, which as theoretically expected does not depend on $\mbox{Ek}$ (figure \ref{fig:Ek}(a)). The remaining difference at small $\mbox{Ek}$ might be attributed to still neglecting curvature effects of the perturbations in the analytical approach.

\begin{figure}
	\centering
	\includegraphics[width=.7\linewidth]{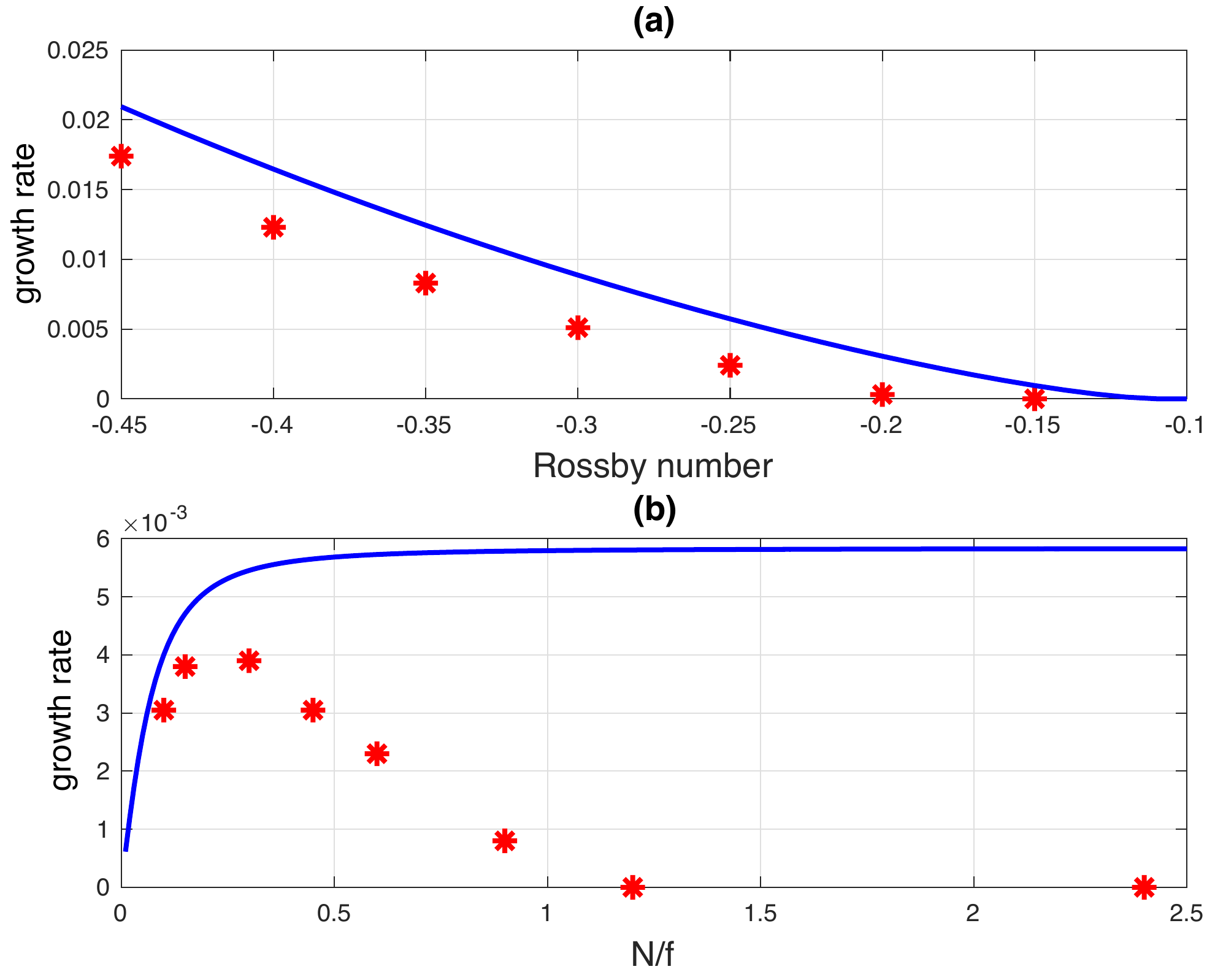}
	\caption{Systematic study of the growth rate as a function of the Rossby number $\mbox{Ro}$ for $\mbox{Sc}=200$, $\mbox{Ek}=3.5\times 10^{-4}$, and $N/f=0.6$ (a) and as a function of the ratio $N/f$ for $\mbox{Sc}=200$, $\mbox{Ek}=3.5\times 10^{-4}$ and $\mbox{Ro}=-0.25$. The blue curves show predictions from the McIntyre linear theory {accounting for the centrifugal terms}, and red stars show numerical results. Symbol size is larger than estimated uncertainties.}
	\label{fig:Ro_Nf}
\end{figure}
 
{These main conclusions are confirmed by a systematic comparison of the predicted and measured growth rates as a function of $Ro$ and $N/f$, shown in figure \ref{fig:Ro_Nf}.}  
A good qualitative agreement is found for the growth rate evolution as a function of the Rossby number, but the analytical approach systematically overestimates {their magnitude} (figure \ref{fig:Ro_Nf}(a)). And while reasonable agreement is found for the lower values of $N/f$, the predicted and measured growth rates significantly depart when $N/f$ increases (figure \ref{fig:Ro_Nf}(b)). This was to be expected from the previous discussion: with the present choice of adimensionalisation, the height of the floating vortex is proportional to $f/N$, while the wavelength of the theoretically most unstable mode is quasi constant when $N/f$ increases. Increasing $N/f$ thus significantly hampers scale separation, hence the validity of the local approach. In the simulation,  vortices {that are too flat} do not support layering formation since they do not sustain sufficiently strong velocity and density gradients at the layer scale to excite the {McIntyre} instability.

In conclusion, this numerical study definitively confirms the McIntyre origin of the layering observed around some laboratory vortices, in particular in \cite{griffiths1981stability}. A dedicated computation reproducing Griffith \& Linden experimental parameters gives a typical growth time of $875$~s and a most unstable wavelength of $0.78$~cm, compatible with their observations. Pushing further this quantitative comparison is irrelevant since (i) figure \ref{fig:intro}(b) shows the non-linear saturation of the instability, with a wavelength potentially different from the most rapidly growing one in the numerics; and (ii) the present model studies the evolution of the vortex shown in figure \ref{fig:intro}(b) ``frozen'' in terms of size and Rossby number, { as well as for a base state  in thermal wind balance approximated by Gaussian profiles in azimuthal velocity and density anomaly. In} the real experiment, this vortex results from a continuous injection, associated to a progressive growth in size and decrease in  Rossby number through time, { as well as to a possibly different base state}.

\section{Application to real meddies}\label{sec:app}

Previous section has shown that the local analysis performs well to predict the most unstable wavelength as well as the growth rate,  especially when scale separation is ensured.  Let's hence investigate whether or not McIntyre instability could participate in explaining layering around meddies, in addition to the already validated double diffusive \cite{radko2017life} and critical layer \cite{hua2013layering} instabilities. Following the data given e.g. in \cite{aubert2012universal}, a typical meddy like Bobby \cite{pingree1993structure} has $L \simeq 27$~km, $f \simeq 8.3 \times 10^{-5}$~rad/s, $\mbox{Ro} \simeq -0.17$, and $N \simeq 2.3 \times 10^{-3}$~rad/s. Let's take into account the meddy internal stratification with $N_c \simeq 1.7 \times 10^{-3}$~rad/s, which as mentioned in section \ref{sec:num}, is not considered for laboratory experiments but simply introduces a factor $1- N_c^2/N^2$ in the density profile. Since layering appears rapidly after meddy formation (see figure \ref{fig:intro}(a)), an upper bound (in absolute value) for the Rossby number $\mbox{Ro} \simeq -0.35$ is also considered \cite{hua2013layering}, which might more properly characterise the meddy at its formation.  { Note that even for this rather large value of $\mbox{Ro}$, we model here the meddy as a vortex in thermal wind balance with Gaussian profiles of density anomaly and azimuthal velocity. Accounting for more realistic base states could be the subject of feature studies.}

With those parameters and considering molecular values for diffusion, the linear stability analysis predicts that the flow is stable (no layering) when considering temperature stratification with $\mbox{Sc}=7$, while it is unstable (layering formation) when considering salt stratification with $\mbox{Sc}=700$. The predicted growth time and most unstable wavelength are then $114$~days and $0.91$~m when considering $\mbox{Ro} = -0.17$, and $24.0$~days and $0.86$~m when considering $\mbox{Ro} = -0.35$. The linear growth time hence seems rather long and the selected wavelength rather short for anticipating a significant involvment of this mechanism in the global layering pattern shown figure \ref{fig:intro}(a). For comparison, \cite{nguyen2012slow} envisages a growth time at least 3 times smaller and wavelengths 5 to 70 times larger for layering from a critical layer instability. 

These conclusions can nevertheless be challenged when considering turbulent diffusivities that could more relevantly describe oceanic processes at the layer scale in the vicinity of a real meddy side. Typical values for  turbulent viscosity range from $1$ to $10^5$ times the molecular values (see e.g. the discussion in \cite{facchini2016lifetime}), and even more for turbulent heat and salt diffusion (see e.g. \cite{radko2017life}). One could claim that when considering turbulent diffusion,  a turbulent Schmidt number {close to unity} should be accordingly considered {  \cite[see e.g.][]{gualtieri2017values}},  which would preclude McIntyre instability.  But it can also be argued that in the physical mechanism underlying the McIntyre instability, the horizontal momentum diffusion opposes the vertical diffusion of the stratifying agent for sustaining the instability: hence the relevant anisotropic turbulent Schmidt number could be {very} different from $1$. Then of course, this anisotropy would bring new questions regarding e.g. the instability threshold and angle / wavelength selection. Clearly,  {choosing} the relevant values for turbulent viscosity, as well as for turbulent heat and salt diffusion, is a complex, open issue beyond the scope of the present paper. Here, a reciprocal approach is rather proposed, addressing the following question: in the context of a mesoscale modeling of the meddies and according to McIntyre linear theory, what would be the range of turbulent viscosity and isotropic turbulent Schmidt number for expecting a linear growth compatible with layering observations, i.e. a growth time smaller than $\sim 10$ days and a wavelength of $10-100$~m?

\begin{figure}
	\centering
	\includegraphics[width=.8\linewidth]{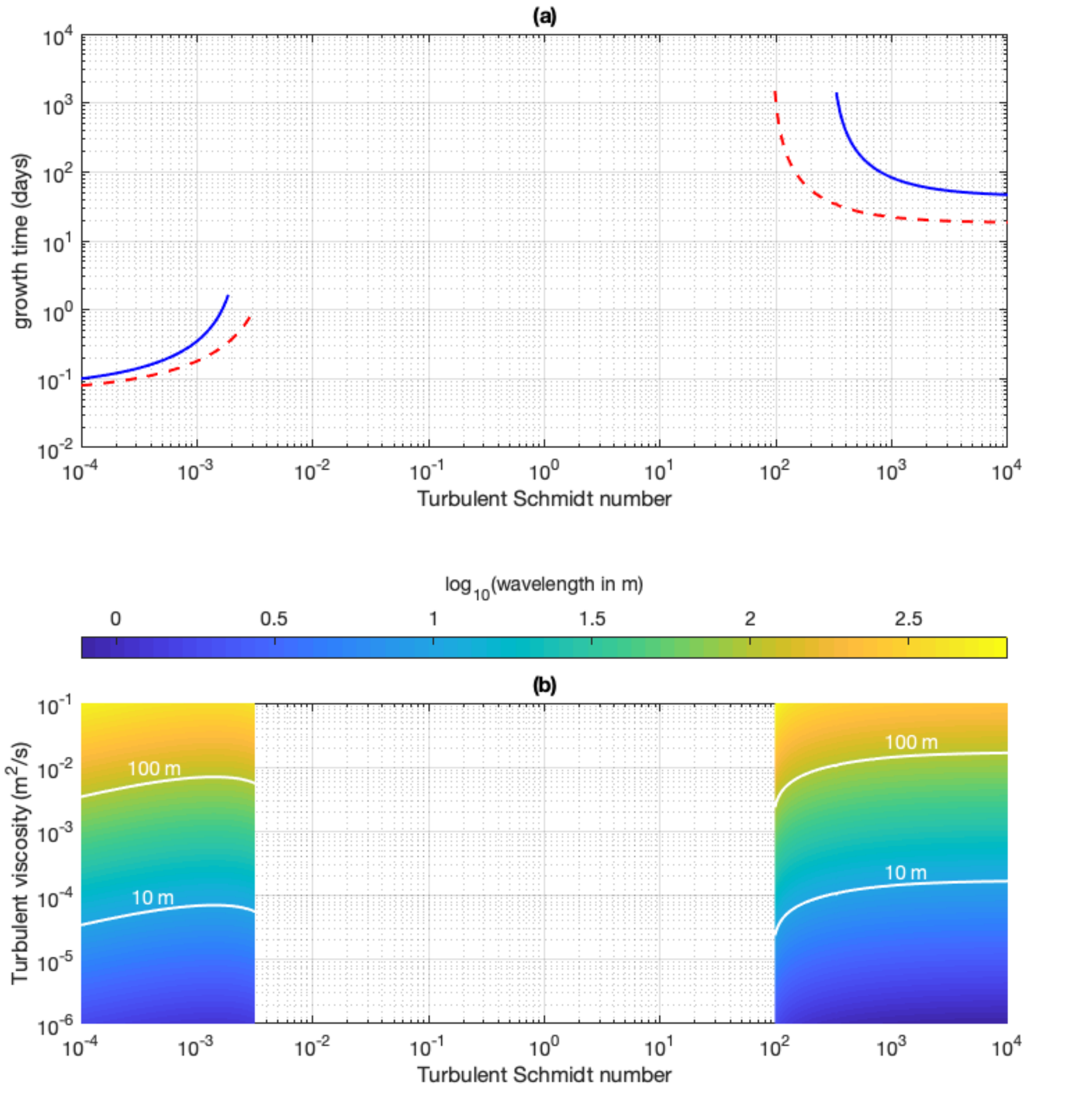}
	\caption{Growth time (a) and wavelength of the most unstable mode (b) for the McIntyre instability of a typical meddy, for various values of the turbulent Schmidt number and, in (b), for various values of the turbulent viscosity (which does not affect the growth time as seen theoretically in section \ref{sec:local}). The continous blue and dotted red lines in (a) consider respectively $\mbox{Ro}=-0.17$ and $\mbox{Ro}=-0.35$. (b) considers the case $\mbox{Ro}=-0.35$, and the white lines highlight iso-values $10$~m and $100$~m.}
	\label{fig:application}
\end{figure}

Results are presented in figure \ref{fig:application}. Only turbulent Schmidt numbers smaller than one can lead to sufficiently fast growing layering around real meddies, typically $\mbox{Sc} \leq 3.2\times 10^{-3}$ for a growth time below a day with the most extreme $\mbox{Ro}$. Then, relevant layer sizes can be produced by a turbulent viscosity of order $5\times 10^{-5} - 5\times 10^{-3}$~m$^2$/s, corresponding to turbulent diffusivities of order $1.6\times 10^{-2} - 1.6$~m$^2$/s for the least demanding $\mbox{Sc} = 3.2\times 10^{-3}$. While not completely impossible, those values are at the edge of the typical ranges considered for oceanic applications \cite[e.g.][]{garabato2004widespread,radko2017life}. It should be noticed that the Gaussian model used here as the base flow leads to rather soft gradients of vorticity and density, which temper the development of the {McIntyre} instability. Rather than using a global, meso-scale representation of the full floating vortex, it would be interesting in a dedicated study to re-investigate the question using local values of the real oceanic base flow measured before the development of any instability and associated signal blurring. Note however that the seminal study \cite{ruddick1992intrusive} also dismissed McIntyre instability as a plausible cause for meddy layering on the basis of such a local approach, considering the observed layers angle with the vertical.

\section{Conclusion and future work}\label{sec:conclusion}

In conclusion, this study based on the systematic exploration of a simple Gaussian-like configuration using axisymmetric numerical simulations has proven that the McIntyre instability is responsible for the layering observed around laboratory vortices, and that the local analysis first proposed in \cite{mcintyre1970diffusive} explains reasonably well the observed growth rates and wavelengths,  {especially} when scale-separation between the base flow and the excited modes is ensured. Extending this local approach towards oceanic typical values does not support the relevance of the {McIntyre} instability for contributing to the layering observed around meddies, but cannot completely dismiss it neither. 

Several directions can be envisaged to extend these conclusions. From the experimental point of view, since purely fluid, floating vortices experiments can only explore a very limited range of parameters, one could think of extending the historical spin-up experiments \cite{baker1971density,calman1977experiments}: anticyclonically rotating a solid ellipsoid with an aspect ratio equal to the equilibrium shape of floating vortices given by (\ref{eq:aspect}) would generate a flow close to the meddy configuration (then assumed to be in solid body rotation) and would allow to explore an extended range of Rossby number. Changing the aspect ratio would allow to study the competition between the critical layer instability and the McIntyre instability \cite{meunier2014instabilities}. Besides, thermalising the whole set-up or using different salts would allow to explore an extended range of Schmidt number. Combining salt stratification and thermal effects (by heating the ellipsoid) would allow to also explore the competition with the double diffusive instability, as relevant for meddies. From the numerical point of view, it is obviously necessary now to explore three-dimensional and turbulent flows, accounting both for the small Ekman number of real meddies and for their chaotic environment: this is already done in astrophysics in the $\mbox{Sc}\ll 1$ limit \cite{barker2020angular}, which actually could also be relevant for oceanic applications (see section \ref{sec:app}). Beyond this {computational} challenge, it would already be interesting, with the present basic numerical tool, to study the non-linear, longterm evolution of the instability in order to quantify its saturation state and its mixing efficiency.  {It would also be interesting to consider other types of base flow, including gradient wind balanced ones.} All these points will be the subject of future works.

\section*{Acknowledgment}
The author acknowledges funding by the European Research Council under the European Union's Horizon 2020 research and innovation program through Grant No. 681835-FLUDYCO-ERC-2015-CoG. 
The results presented here have beneficiated from preliminary studies performed during the Ph.D of Oriane Aubert \cite{aubert2013formes}, as well as from fruitful discussions with Patrice Le Gal and Daphn\'e Lemasquerier.
\bibliographystyle{ieeetr}

\end{document}